\documentclass[11pt,a4paper,reqno]{article}
\usepackage{mysty}
\usepackage{authblk}

\usepackage{fullpage}
\usepackage{subfig}

\usepackage{amsfonts,amssymb}
\usepackage{amsmath}
\usepackage{graphicx}
\usepackage{cite}
\usepackage{enumerate}

\theoremstyle{plain}

\usepackage{babel}
\providecommand{\theoremname}{Theorem}

\SetLabelAlign{parright}{\parbox[t]{\labelwidth}{\raggedleft{#1}}}
\setlist[description]{style=multiline,topsep=4pt,align=parright}%,font=\normalfont

\makeatletter
\let\reftagform@=\tagform@
\def\tagform@#1{\maketag@@@{(\ignorespaces\textcolor{black}{#1}\unskip\@@italiccorr)}}
\newcommand{\iref}[1]{\textup{\reftagform@{\tcr{\ref{#1}}}}}
\makeatother

\begin{document}
	
%% TITLE %%%%%%%%%%%%%%%%%%%%%%%%%%%%%%%%%%
\title{Library-based Fast Algorithm for Simulating the Hodgkin-Huxley Neuronal Networks}
\author{Zhong-Qi Kyle Tian and Douglas Zhou\textsuperscript{\footnote{zdz@sjtu.edu.cn}}}
\affil{School of Mathematical Sciences, MOE-LSC, and Institute of Natural Sciences, Shanghai Jiao Tong University, Shanghai, China}

\date{}
\maketitle
%\begin{affiliations}
%	\item School of Mathematical Sciences, MOE-LSC, and Institute of Natural Sciences, Shanghai Jiao Tong University, Shanghai, China	
%\end{affiliations}

%%%%%%%%%%%%%%%%%%%%%%%%%%%%%%%%%%%
\begin{abstract}
We present a modified library-based method for simulating the Hodgkin-Huxley
(HH) neuronal networks. By pre-computing a high resolution data library
during the interval of an action potential (spike), we can avoid evolving
the HH equations during the spike and can use a large time step to
raise efficiency. The library method can stably achieve at most 10
times of speedup compared with the regular Runge-Kutta method while
capturing most statistical properties of HH neurons like the distribution
of spikes which data is widely used in the statistical analysis like
transfer entropy and Granger causality. The idea of library method
can be easily and successfully applied to other HH-type models like
the most prominent \textquotedblleft regular spiking\textquotedblright ,
\textquotedblleft fast spiking\textquotedblright , \textquotedblleft intrinsically
bursting\textquotedblright{} and \textquotedblleft low-threshold spike\textquotedblright{}
types of HH models. 
	
\end{abstract}

\section{Introduction}

The Hodgkin-Huxley (HH) model \cite{hodgkin1952quantitative,hassard1978bifurcation,dayan2003theoretical},
originally proposed to describe the behavior of the giant axon, is
one of the most realistic models. It is then regarded as the foundation
for other neuronal models with more complicated behaviors like bursting
and adaptation. Because of its complexity, we often use regular Runge-Kutta
scheme in the numerical simulation to study the dynamics of this model.
However the HH equations become stiff when the neuron fires a spike.
As a consequence, we have to take a sufficiently small time step to
avoid stability problems. But we need to simulate the model quite
frequently to study its properties with different aims. Sometimes
we even need to simulate the model with hundreds of neurons for hours
to record the spikes or voltages \cite{ito2011extending,zhou2014granger}.
Therefore, it is important to find a fast algorithm to simulate the
model. 

The stiff part during a firing event is from the activities of its
sodium and potassium ion channels and can last for about 3 ms. We offer
a library method to deal with the stiff period and it can use much
larger time steps compared with the regular Runge-Kutta method. The
library method treats a HH neuron as an I\&F one. Once a HH neuron's
membrane potential reaches the threshold, we stop evolving its HH
equations and restart after the stiff part. The time-courses of membrane
potential and gating variables during the stiff part can be recovered
from a pre-computed high resolution data library. So once the membrane
potential reaches the threshold, we record its state and decide the
restart state interpolated from the data library. Therefore, we can
avoid the stiff part and use a large time step to evolve the HH model.
The library method can use time steps one order of magnitude larger
than the regular Runge-Kutta method while achieving precise statistical
information of the HH model, $e.g.$, the distribution of spikes.
Recently, statistical tools like Granger causality, transfer entropy
and maximum entropy have been proven to be effective in probing neural
interactions, $e.g.$, detecting causality, identifying effective
connectivity and reconstructing the fire patterns \cite{ito2011extending,zhou2013causal,zhou2014granger,xu2017dynamical}.
These works are mainly based on the spike trains recorded for $\sim20$
minutes to obtain an accurate distribution. We specially point out
that the library method can stably speed up at most 10 times compared
with the regular Runge-Kutta methods, which may make the HH model
attractive as a base model in these statistical tools. 

We emphasize that we should take into account the causality of synaptic
spikes within a single time step. In general, when we evolve the HH
neurons for one time step in the regular Runge-Kutta methods \cite{hansel1998numerical,shelley2001efficient},
we only use the feedforward input during the time step. Without knowing
when and which neurons will fire during the time step, we have to
wait until the end of the evolution of this time step to consider
the effect of these possible synaptic spikes. This approach may work
well with a sufficiently small time step that there are only $O(1)$
spikes during one time step and the causal effect is negligible. However,
when we use a large time step, the first spikes may influence the
network via the spike-spike interactions that some extra spikes may
appear if the first spikes are excitatory and some latter spikes may
vanish otherwise. Therefore, to use a large time step, we should take
the spike-spike correction procedure \cite{rangan2007fast} to obtain
accurate spike sequences. 

We also investigate the validity of the library method in more complicated
HH-type models with more voltage-dependent currents. They are the
four most prominent types: \textquotedblleft fast spiking\textquotedblright ,
\textquotedblleft regular spiking\textquotedblright , \textquotedblleft intrinsically
bursting\textquotedblright{} and \textquotedblleft low-threshold spike\textquotedblright{}
\cite{izhikevich2003simple,pospischil2008minimal}. For each type
of model, we build and use the data library in the same way as that
in the standard HH model. The library method can still achieve accurate
statistical information of these HH-type models with remarkable computational
speedup.

\section{Materials and methods}

\subsection{HH model \label{sec:The-model}}

The dynamics of the $i$th neuron of an excitatory Hodgkin-Huxley
(HH) neuronal network is governed by 

\begin{equation}
\begin{cases}
\begin{aligned}C\frac{dV_{i}}{dt} & =-(V_{i}-V_{\textrm{Na}})G_{\textrm{Na}}m_{i}^{3}h_{i}-(V_{i}-V_{\textrm{K}})G_{\textrm{K}}n_{i}^{4}-(V_{i}-V_{\textrm{L}})G_{\textrm{L}}+I_{i}^{\textrm{input}}\\
\frac{dm_{i}}{dt} & =(1-m_{i})\alpha_{m}(V_{i})-m_{i}\beta_{m}(V_{i})\\
\frac{dhi}{dt} & =(1-h_{i})\alpha_{h}(V_{i})-h_{i}\beta_{h}(V_{i})\\
\frac{dn_{i}}{dt} & =(1-n_{i})\alpha_{n}(V_{i})-n_{i}\beta_{n}(V_{i})
\end{aligned}
\end{cases}\label{eq: HH}
\end{equation}
where $V_{i}$ is the membrane potential, $m_{i}$, $h_{i}$ and $n_{i}$
are gating variables, $I_{i}^{\textrm{input}}$ is the input current,
$\alpha$ and $\beta$ are empirical functions of $V$,

\begin{equation}
\begin{aligned} & \alpha_{m}(V_{i})=\frac{0.1V_{i}+4}{1-\exp(-0.1V_{i}-4)} &  & \beta_{m}(V_{i})=4\exp(-(V_{i}+65)/18)\\
& \alpha_{h}(V_{i})=0.07\exp(-(V_{i}+65)/20) &  & \beta_{h}(V_{i})=\frac{1}{1+\exp(-3.5-0.1V_{i})}\\
& \alpha_{n}(V_{i})=\frac{0.01V_{i}+0.55}{1-\exp(-0.1V_{i}-5.5)} &  & \beta_{n}(V_{i})=0.125\exp(-(V_{i}+65)/80)
\end{aligned}
\end{equation}
Other parameters are constants: $C=1\mu\textrm{F\ensuremath{\cdot}cm}^{-2}$
is the membrane capacitance; $V_{\textrm{Na}}=50$ mV, $V_{\textrm{K}}=-77$
mV and $V_{\textrm{L}}=-54.387$ mV are reversal potentials; $G_{\textrm{Na}}=120\textrm{ mS\ensuremath{\cdot}cm}^{-2}$,
$G_{\textrm{K}}=36\textrm{ mS\ensuremath{\cdot}cm}^{-2}$, and $G_{\textrm{L}}=0.3\textrm{ mS\ensuremath{\cdot}cm}^{-2}$
are the maximum conductances. 

The input current $I_{i}^{\textrm{input}}$ is given by $I_{i}^{\textrm{input}}=-G_{i}(t)(V_{i}-V_{G})$
with

\begin{equation}
\frac{dG_{i}(t)}{dt}=-\frac{G_{i}(t)}{\sigma_{\textrm{r}}}+H_{i}(t)
\end{equation}

\begin{equation}
\frac{dH_{i}(t)}{dt}=-\frac{H_{i}(t)}{\sigma_{\textrm{d}}}+f\sum_{l}\delta(t-s_{il})+\sum_{j\neq i}\sum_{l}S_{ij}\delta(t-\tau_{jl})\label{eq:f input}
\end{equation}
where $V_{\textrm{G}}$ is the reversal potential with value $V_{\textrm{G}}=0$
mV, $G_{i}(t)$ is the conductance, $H_{i}(t)$ is an additional parameter
to describe $G_{i}(t)$, $\sigma_{\textrm{r}}$ and $\sigma_{\textrm{d}}$
are fast rise and slow decay time scale, respectively, and $\delta(\cdot)$
is the Dirac delta function. In this Letter, we use $\sigma_{\textrm{r}}=0.5$
ms, $\sigma_{\textrm{d}}=3.0$ ms. The second term in Eq. (\ref{eq:f input})
is the feedforward input with magnitude $f$. The input time $s_{il}$
is generated from a Poisson process with rate $\nu$. The third term
in Eq. (\ref{eq:f input}) is the synaptic current from synaptic interactions
in the network, where $S_{ij}$ is the coupling strength from the
$j$th neuron to the $i$th neuron, $\tau_{jl}$ is the $l$th spike
time of $j$th neuron. 

When the voltage $V_{i}$, evolving continuously according to Eq.
(\ref{eq: HH}), reaches the threshold $V^{\textrm{th}}$, we say
the $i$th neuron fires a spike at this time. Instantaneously, all
its postsynaptic neurons receive this spike and their corresponding
parameter $H$ jumps by an appropriate amount, $\mathit{e.g.}$, $S_{ji}$
for the $j$th neuron. For the sake of simplicity, we mainly consider
an all-to-all coupled network with $S_{ij}=S/N$, where $S$ is the
coupling strength and $N$ is the number of neurons. The given methods
can be easily extended to other types of networks, $\mathit{e.g.}$,
with inhibitory neurons, randomly and inhomogeneously connected.

\subsection{Numerical scheme}

We first introduce the most widely used Runge-Kutta fourth-order scheme
(RK4) with fixed time step $\Delta$t to evolve the HH model. Since
the neurons interact with each other through the spikes by influencing
of conductance of the postsynaptic neurons, it is important to obtain
accurate spike sequences. Then, there are some issues need to be clarified
in simulation. For example, how to determine the spike time accurately
\cite{hansel1998numerical,shelley2001efficient}. Suppose the $i$th
neuron fires a spike at time $\tilde{t}$ in $[t,t+\Delta t]$, a
naive way to determine the spike time is to set $\tilde{t}=t+\Delta t$
and then an error of order $\Delta t$ is introduced. Therefore, the
whole scheme is limited to the first-order. To solve this problem,
we can use numerical interpolation schemes to decide the spike time
more accurately \cite{hansel1998numerical,shelley2001efficient}.
After evolving the trajectory of the $i$th neuron from $t$ to $t+\Delta t$,
we can use the obtained values $V_{i}(t),V_{i}(t+\Delta t),\frac{dV_{i}}{dt}(t),\frac{dV_{i}}{dt}(t+\Delta t)$
to perform a cubic Hermite interpolation to decide the spike time.
Then the whole scheme have an accuracy of fourth-order. 

Another problem is the causality of the spike events \cite{rangan2007fast}.
A usually used strategy is to evolve the network (\ref{eq: HH}),
for example from $t$ to $t+\Delta t$, only considering the feedforward
input within the time interval $[t,t+\Delta t]$. If some synaptic
spikes are fired during this interval, they will be assigned at the
end of the time step $t+\Delta t$. This strategy will lead to some
problems. One is that since we assign the synaptic spikes at the end
of the time step rather than the real spike times, the accuracy is
limited to the first-order. Another problem is that the first few
synaptic spikes may strongly influence the other spiking neurons by
spike-spike interactions, especially in the simulation with a large
time step, hence the rest of the synaptic spikes may be incorrect.

To solve this problem, we take the spike-spike correction procedure
\cite{rangan2007fast}, which strategy is similar to the event-driven
approach \cite{mattia2000efficient,reutimann2003event,rudolph2007much}.
Suppose we evolve the all-to-all connected network from $t$ to $t+\Delta t$.
Here is the details. We first preliminarily evolve the neurons in
the network independently from $t$ to $t+\Delta t$ considering only
the feedforward input. If any neuron fires a spike during this time
interval, say the $i$th neuron, we denote the spike time by $t_{\textrm{fire}}^{(i)}$,
along with the cubic Hermite interpolation to determine the spike
time. If a neuron does not fire during $[t,t+\Delta t]$, still say
the $i$-th neuorn, we then set $t_{\textrm{fire}}^{(i)}=t+\Delta t$.
After obtaining all these $t_{\textrm{fire}}^{(i)}$ values independently,
we find the minimum one, without loss of generality, say $t_{\textrm{fire}}^{(1)}$.
If $t_{\textrm{fire}}^{(1)}=t+\Delta t$, then there are no synaptic
spikes in $[t,t+\Delta t]$ and therefore, there is no causal problem
and we accept the preliminary trajectories as the solution. Otherwise,
$t_{\textrm{fire}}^{(1)}$ is the first synaptic spike in $[t,t+\Delta t]$
and there is no causal problem during $[t,t_{\textrm{fire}}^{(1)}]$.
So we update all the neurons from $t$ to $t_{\textrm{fire}}^{(1)}$,
at which time neuron 1 still has a firing event. Then we move on to
start another loop to find the next first synaptic spike time in $[t_{\textrm{fire}}^{(1)},t+\Delta t]$
until the evolving is finished. 

With the cubic Hermite interpolation and spike-spike correction, we
give the regular RK4 scheme. For the easy of illustration, we use
the vector 
\begin{equation}
X_{i}(t)=(V_{i}(t),m_{i}(t),h_{i}(t),n_{i}(t),G_{i}(t),H_{i}(t))\label{eq:stats X}
\end{equation}
to represent the state of the $i$th neuron. Details of the numerical
algorithm to evolve the network from $t$ to $t+\Delta t$ is given
in Algorithm 1.

\begin{algorithm}[H] \label{Algo: 1}  	  
	% no vertical line  
	%\SetAlgoNoLine  	 
	\caption{Regular RK4 algorithm} 
	\KwIn{$t$, $\Delta t$, $\{X_i(t)\}$ and $\{s_{il}\}$} 
	\KwOut{$\{X_i(t+\Delta t)\}$ and $\{\tau_{il}\}$(if any fired)} 
	Preliminarily evolve the network from $t$ to $t+\Delta t$ to find the first synaptic spike: 
	
	\For{$i = 1$ to $N$} 
	{Let $M$ denote the total number of feedforward spikes of the $i$th neuron within $[t, t+\Delta t]$ and sort them into an increasing list $\{T_m^{\text{sorted}}\}$. Then we extend this notation such that 		$T_0^{\text{sorted}} = t$ and $T_{M+1}^{\text{sorted}} = t+\Delta t$.\\
		
		\For{$m = 1$ to $M+1$ } 	
		{Advance the equations for the $i$th HH neuron from  $T_{m-1}^{\text{sorted}}$ to $T_m^{\text{sorted}}$ using the standard RK4 scheme. Then update the conductance $H_i(T_m^{\text{sorted}})$ by adding $f$.\\				 		
			\eIf{$V_i(T_{m-1}^{\text{sorted}}) < V^{\textrm{th}}$, $V_i(T_m^{\text{sorted}}) \geq V^{\textrm{th}}$} 		
			{Find the spike time $t_{\text{fire}}^{(i)}$ by cubic Hermite interpolation using the values $V_i(T_{m-1}^{\text{sorted}})$, $V_i(T_m^{\text{sorted}})$, 				$\frac{dV_i}{dt}(T_{m-1}^{\text{sorted}})$ and $\frac{dV_i}{dt}(T_m^{\text{sorted}})$.\\ } 		
			{Set $t_{\text{fire}}^{(i)} = t+\Delta t$. } 	
		} 
	}	 
	\While{The minimum of $\{t_{\text{fire}}^{(i)}\} < t+\Delta t$ } 
	{Suppose  $t_{\text{fire}}^{(1)}$ is the minimum one.\\ 		Evolve neuron 1 to $t_{\text{fire}}^{(1)}$, and generate a spike at this moment. Then update all the remaining neurons to  $t_{\text{fire}}^{(1)}$.\\							
		Preliminarily evolve the network from $t_{\text{fire}}^{(1)}$ to $t+\Delta t$ to find the next first synaptic spike.									 	}			 
	{We accept $\{X_i{(T_{M+1}^{\text{sorted}})}\}$ as the solution  $\{X_i{(t+\Delta t)}\}$;\\ 	} 
\end{algorithm}

\subsection{Library method}

\begin{figure}[H]
	\begin{centering}
		\includegraphics[width=1.0\textwidth ]{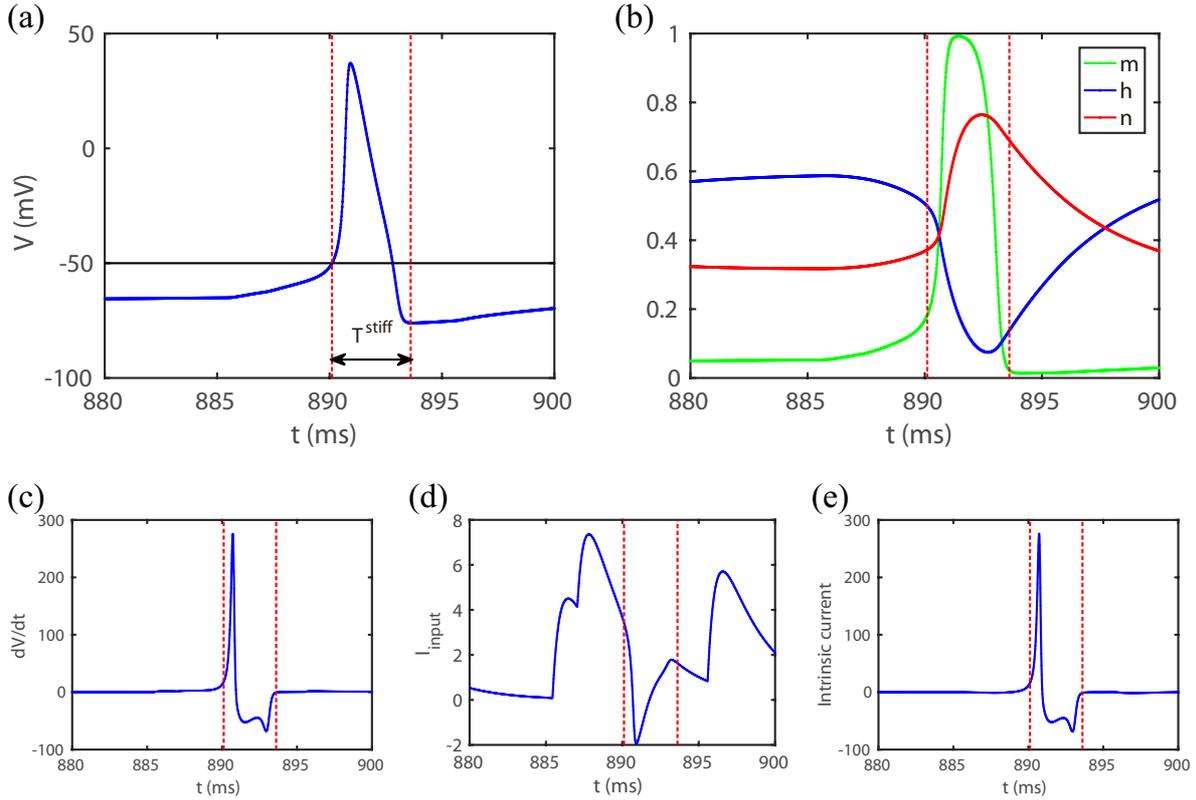}
		\par\end{centering}
	\caption{Typical firing event. (a) The trajectory of voltage $V$. The black
		line indicates the threshold $V^{\textrm{th}}$ and the red dotted
		lines indicate the stiff period. (b) The trajectory of $m,h,n$. (c)
		The trajectory of $\frac{dV}{dt}$ . (d) The trajectory of the input
		current $I_{i}^{\textrm{input}}$ ($\mu\textrm{A\ensuremath{\cdot}cm}^{-2}$).
		(e) The trajectory of the intrinsic current ($\mu\textrm{A\ensuremath{\cdot}cm}^{-2}$)
		which is the sum of ionic and leakage currents. \label{fig:Typical-fire-pattern}}
\end{figure}

When a neuron fires a spike, the HH neuron equations are stiff for
some milliseconds denoted by $T^{\textrm{stiff}}$, as shown in Fig
\ref{fig:Typical-fire-pattern}. This stiff period requires a sufficiently
small time step to avoid stability problem. Therefore, in the regular
RK4 scheme, we have to use a relatively small time step, $\mathit{e.g.}$,
the widely used $\Delta t=1/32$ ms. To overcome the limitation in
time step, we propose a modified library method \cite{sun2009library}.
It is based on the regular RK4 scheme and has the advantage of using
a large time step to raise efficiency, while having comparable accuracy
in statistical quantifications, $\mathit{e.g.}$, mean firing rate
and the spike pattern (Details are given in Section \ref{sec: Numerical result}).
The library method depends on the length of stiff period, which is
experientially set $T^{\textrm{stiff}}=3.5$ ms, long enough to cover
the stiff parts in general firing events. 

The library method \cite{sun2009library} treats the HH neuron's firing
event like an I\&F neuron. Once a neuron's membrane potential reaches
the threshold $V^{\textrm{th}}$, its voltage rises and reaches the
peak value very quickly because of a large influx of the sodium current,
then it drops back down to the lowest point by the potassium current.
This process is actually an action potential and lasts for about 3
ms which is indeed the stiff period, as shown in Fig \ref{fig:Typical-fire-pattern}(a).
If we have a pre-computed high resolution data library of $V,m,h,n$,
we can recover their time-courses. In other words, once a neuron's
membrane potential reaches the threshold $V^{\textrm{th}}$, we stop
evolving $V,m,h,n$ for the following stiff period $T^{\textrm{stiff}}$,
and restart with the values interpolated from the library. Thus the
stiff part is avoided and we can use a large time step to evolve the
model to raise efficiency. 

\subsubsection{Build the library}

Now we describe how to build the library in detail. Once a neuron's
membrane potential reaches the threshold $V^{\textrm{th}}$, we record
the values $I^{\textrm{input}},m,h,n$ and denote them by $I^{\textrm{th}},m^{\textrm{th}},h^{\textrm{th}},n^{\textrm{th}}$,
respectively. If we know the exact trajectory of $I^{\textrm{input}}$
for the following stiff period $T^{\textrm{stiff}}$, we can use a
sufficiently small time step to evolve the Eq. (\ref{eq: HH}) for
$T^{\textrm{stiff}}$ with initial values $V^{\textrm{th}},m^{\textrm{th}},h^{\textrm{th}},n^{\textrm{th}}$
to obtain high resolution trajectories of $V,m,h,n$. We denote the
obtained values after evolving by $V^{\textrm{re}},m^{\textrm{re}},h^{\textrm{re}},n^{\textrm{re}}$,
where the superscript $\textrm{-re}$ stands for the reset value. 

However it is impossible to obtain the exact trajectory of $I^{\textrm{input}}$
without knowing the feedforward and synaptic spike information. As
shown in Fig \ref{fig:Typical-fire-pattern}(d, e), $I^{\textrm{input}}$
varies during the stiff period with peak value in the range of $O(5)$
$\mu\textrm{A\ensuremath{\cdot}cm}^{-2}$, while the intrinsic current,
the sum of ionic and leakage current, is about 30 $\mu\textrm{A\ensuremath{\cdot}cm}^{-2}$
at the spike time, and quickly rises to the peak value about 250 $\mu\textrm{A\ensuremath{\cdot}cm}^{-2}$,
then stays at $O(-50)$ $\mu\textrm{A\ensuremath{\cdot}cm}^{-2}$
in the remaining stiff period. Therefore, the intrinsic current is
dominant in the stiff period. With this observation, we keep $I^{\textrm{input}}$
as constant throughout the whole stiff period. We emphasize that this
is the only assumption made in the library method. Then, given a suite
of $I^{\textrm{th}},m^{\textrm{th}},h^{\textrm{th}},n^{\textrm{th}}$,
we can obtain a suite of $V^{\textrm{re}},m^{\textrm{re}},h^{\textrm{re}},n^{\textrm{re}}$. 

\begin{figure}[H]
	\begin{centering}
		\includegraphics[width=0.75\textwidth]{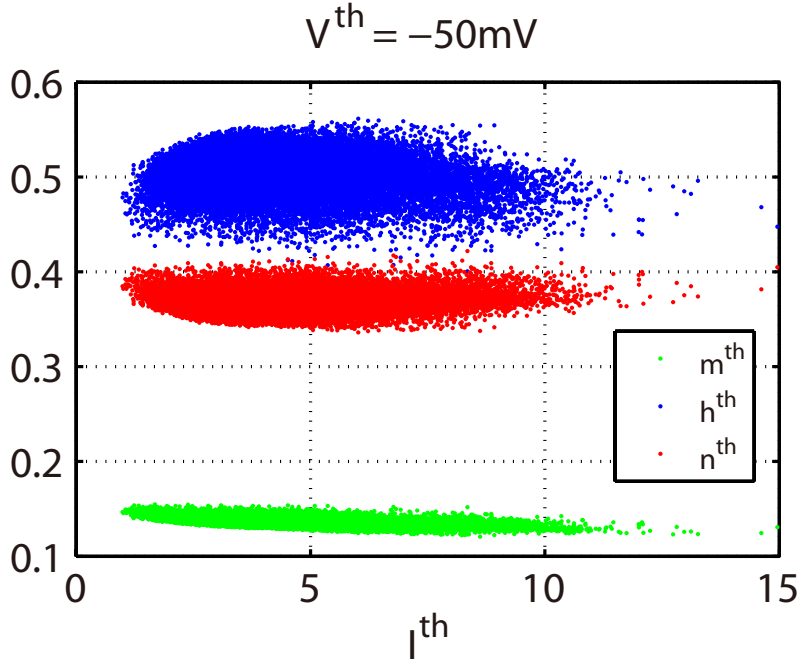}
		\par\end{centering}
	\caption{The ranges of values $I^{\textrm{th}},m^{\textrm{th}},h^{\textrm{th}},n^{\textrm{th}}$.
		\label{fig:Imhn range} }
\end{figure}

Before building the library, we choose $N_{I},N_{m},N_{h},N_{n}$
different values of $I^{\textrm{th}},m^{\textrm{th}},h^{\textrm{th}},n^{\textrm{th}}$,
respectively, equally distributed in their ranges as shown in Fig
\ref{fig:Imhn range}.  For each suite of $I^{\textrm{th}},m^{\textrm{th}},h^{\textrm{th}},n^{\textrm{th}}$,
we evolve the Eq. (\ref{eq: HH}) for a time interval of $T^{\textrm{stiff}}$
to obtain $V^{\textrm{re}},m^{\textrm{re}},h^{\textrm{re}},n^{\textrm{re}}$
with a sufficiently small time step, $\mathit{e.g.}$, $\Delta t=2^{-16}\approx1.52\times10^{-5}$
ms. Note that we should set $I^{\textrm{input}}$ constant $I^{\textrm{input}}=I^{\textrm{th}}$
throughout the whole time interval $T^{\textrm{stiff}}$. Then the
data library is built with total size $8N_{I}N_{m}N_{h}N_{n}$. Larger
values of $N_{I},N_{m},N_{h},N_{n}$ can increase the accuracy of
the library and greatly increase the size of library at the same time.
In our simulation, we take $N_{I}=21,N_{m}=16,N_{h}=21,N_{n}=16$,
which can make the library sufficiently accurate as is shown in section
\ref{sec: Numerical result}. The library occupies only 6.89 megabyte
in binary form and is quite small for today's computers.

One key point in building the library is to choose a proper threshold
value $V^{\textrm{th}}$. The threshold should be relatively low to
keep the HH equations not stiff and allow a large time step with the
stability requirement satisfied. On the other hand, it should be relatively
high that a neuron will definitely fire a spike after its membrane
potential reaches the threshold. In this Letter, we take $V^{\textrm{th}}=-50$
mV. 

\subsubsection{Use the library}

We now illustrate how to use the library. Once a neuron's membrane
potential reaches the threshold, we first record the values $I^{\textrm{th}},m^{\textrm{th}},h^{\textrm{th}},n^{\textrm{th}}$,
then stop evolving its HH equations of $V,m,h,n$ for the following
$T^{\textrm{stiff}}$ ms and restart with values $V^{\textrm{re}},m^{\textrm{re}},h^{\textrm{re}},n^{\textrm{re}}$
linearly interpolated from the pre-computed high resolution data library.
For the easy of writing, suppose $I^{\textrm{th}}$ falls between
two data points $I_{0}^{\textrm{th}}$ and $I_{1}^{\textrm{th}}$
in the library. Simultaneously find the data points $m_{0}^{\textrm{th}}$
and $m_{1}^{\textrm{th}}$, $h_{0}^{\textrm{th}}$ and $h_{1}^{\textrm{th}}$,
$n_{0}^{\textrm{th}}$ and $n_{1}^{\textrm{th}}$, respectively. So
we need 16 suites of values in the library to do a linear interpolation.
Then the linear interpolation for $V^{\textrm{re}}$ is 

\begin{equation}
V^{\textrm{re}}=\sum_{i,j,k,l\in\{0,1\}}V^{\textrm{re}}(I_{i}^{\textrm{th}},m_{j}^{\textrm{th}},h_{k}^{\textrm{th}},n_{l}^{\textrm{th}})\frac{I^{\textrm{th}}-I_{1-i}^{\textrm{th}}}{I_{i}^{\textrm{th}}-I_{1-i}^{\textrm{th}}}\frac{m^{\textrm{th}}-m_{1-j}^{\textrm{th}}}{m_{j}^{\textrm{th}}-m_{1-j}^{\textrm{th}}}\frac{h^{\textrm{th}}-h_{1-k}^{\textrm{th}}}{h_{k}^{\textrm{th}}-h_{1-k}^{\textrm{th}}}\frac{n^{\textrm{th}}-n_{1-l}^{\textrm{th}}}{n_{l}^{\textrm{th}}-n_{1-l}^{\textrm{th}}}\label{eq:Lib-V}
\end{equation}
Same results hold for the computing of $m^{\textrm{re}},h^{\textrm{re}},n^{\textrm{re}}$. 

As for the parameters $G$ and $H$, obviously, they are not affected
by $V,m,h,n$ and are evolved as usual. After obtaining the high resolution
library, we can use a large time step to evolve the HH neuron network.
The detailed numerical algorithm is given in Algorithm 2.

\begin{algorithm}[H] \label{Algo: 2} 
	% no vertical line 	%\SetAlgoNoLine 
	\caption{Library algorithm} 	
	\KwIn{$t$, $\Delta t$, $\{X_i(t)\}$ and $\{s_{il}\}$} 
	\KwOut{$\{X_i(t+\Delta t)\}$ and $\{\tau_{il}\}$(if any fired)} 
	Preliminarily evolve the network from $t$ to $t+\Delta t$ to find the first synaptic spike: 
	
	\For{$i = 1$ to $N$} 
	{Let $M$ denote the total number of feedforward spikes of the $i$th neuron within $[t, t+\Delta t]$ and sort them into an increasing list $T_m^{\text{sorted}}$. Then we extend this notation such that 		$T_0^{\text{sorted}} = t$ and $T_{M+1}^{\text{sorted}} = t+\Delta t$.\\	 	
		\For{$m = 1$ to $M+1$ } 	
		{Advance the equations for the $i$th HH neuron from  $T_{m-1}^{\text{sorted}}$ to $T_m^{\text{sorted}}$ using the standard RK4 scheme. Then update the conductance $H_i(T_m^{\text{sorted}})$ by adding $f$ .\\				 		
			\eIf{$V_i(T_{m-1}^{\text{sorted}}) < V_{th}$, $V_i(T_m^{\text{sorted}}) \geq V_{th}$} 			
			{Find the spike time $t_{\text{fire}}^{(i)}$ by cubic Hermite interpolation using the values $V_i(T_{m-1}^{\text{sorted}})$, $V_i(T_m^{\text{sorted}})$, 				$\frac{dV_i}{dt}(T_{m-1}^{\text{sorted}})$ and $\frac{dV_i}{dt}(T_m^{\text{sorted}})$.\\ 	 			} 
			{Set $t_{\text{fire}}^{(i)} = t+\Delta t$.}
		} 
	}	 
	\While{The minimum of $\{t_{\text{fire}}^{(i)}\} < t+\Delta t$ } 
	{Suppose  $t_{\text{fire}}^{(1)}$ is the minimum one.\\ 		Evolve neuron 1  to $t_{\text{fire}}^{(1)}$,  and generate a spike at this moment.\\ 	
		Record the values $I^{\textbf{th}}, m^{\textbf{th}}, h^{\textbf{th}}, n^{\textbf{th}}$, then perform a linear interpolation from the library to get $V^{\textbf{re}}, m^{\textbf{re}}, h^{\textbf{re}}, n^{\textbf{re}}$. \\ 	
		Meanwhile, we stop evolving $V, m, h , n$ of neuron 1 for the next $T^{\textbf{stiff}}$ ms, but we still evolve the conductance parameters $G, H$ as usual.\\ 	
		Update all the remaining neurons.\\ 																	
		
		Preliminarily evolve the network from $t_{\text{fire}}^{(1)}$ to $t+\Delta t$ to find the next first synaptic spike.										 	}			 
	{We accept $\{X_i{(T_{M+1}^{\text{sorted}})}\}$ as the solution  $\{X_i{(t+\Delta t)}\}$.\\ 	} 
\end{algorithm}

\subsubsection{Hopf bifurcation and transient states}

The introduced library method is intuitive, regarding the reset values
$V,m,h,n$ as a function of the input values $I^{\textrm{th}},m^{\textrm{th}},h^{\textrm{th}},n^{\textrm{th}}$.
When building the library, we require that the ranges of $I^{\textrm{th}},m^{\textrm{th}},h^{\textrm{th}},n^{\textrm{th}}$
can cover almost all the cases in general firing events. Therefore,
given enough cases of $I^{\textrm{th}},m^{\textrm{th}},h^{\textrm{th}},n^{\textrm{th}}$,
the library method can predict the reset values quite accurately.
We now consider the accuracy of library method with the ideal condition:
one single HH neuron driven by constant input $I^{\textrm{input}}$,
$i.e.$, the assumption made in building the library is satisfied.

\begin{figure}[H]
	\begin{centering}
		\includegraphics[width=1\textwidth]{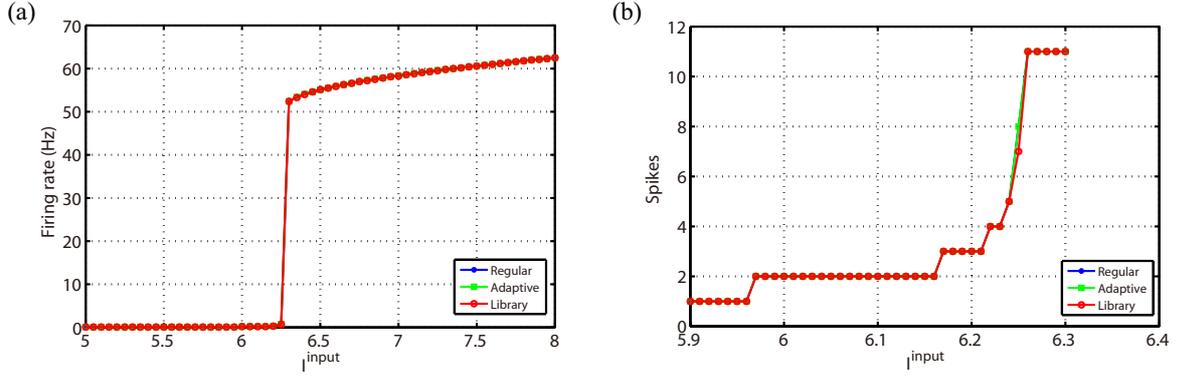}
		\par\end{centering}
	\caption{(a) The firing rate as a function of constant input $I^{\textrm{input}}$
		($\mu\textrm{A\ensuremath{\cdot}cm}^{-2}$). (b) The number of spikes
		during the transient period with initial voltage $V=-65$ mV. The
		blue and red curves in (a) and (b) indicate the regular method with
		$\Delta t=2^{-5}=0.03125$ ms and library method with $\Delta t=0.25$
		ms, respectively. \label{fig:Hopf}}
\end{figure}

There is a type II behavior that only when the input current larger
than a critical value $I^{\textrm{input}}\approx6.2$ $\mu\textrm{A\ensuremath{\cdot}cm}^{-2}$
can a neuron fire regularly and periodically \cite{gerstner2002spiking,koch1998methods}.
The HH model has a sudden jump around this critical value from zero
firing rate to regular nonzero firing rate because of a subcritical
Hopf bifurcation \cite{koch1998methods}, as shown in Fig \ref{fig:Hopf}(a).
Below the critical value, some spikes may appear before the neuron
converges to stable zero firing rate state. The number of spikes during
this transient period depends on how close the constant input is to
the critical value, as shown in Fig \ref{fig:Hopf}(b). Because
our library is built based on the whole information of $I^{\textrm{th}},m^{\textrm{th}},h^{\textrm{th}},n^{\textrm{th}}$,
the library method can indeed capture the Hopf bifurcation and transient
states. We should point out that the library method misses one spike
when $I^{\textrm{input}}=6.25$ $\mu\textrm{A\ensuremath{\cdot}cm}^{-2}$.
This is because the library we use is relatively coarse, with $N_{I}=21,N_{m}=16,N_{h}=21,N_{n}=16$. 

\begin{figure}[H]
	\begin{centering}
		\includegraphics[width=1\textwidth]{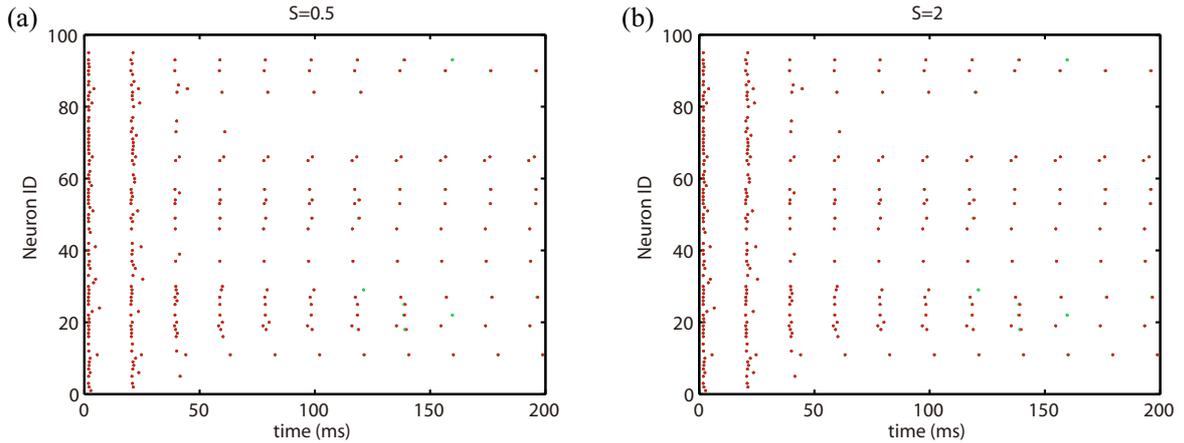}
		\par\end{centering}
	\caption{Raster plots of spike events in an all-to-all connected network with
		100 excitatory neurons, coupling strength (a) $S=0.5$ $\textrm{mS\ensuremath{\cdot}cm}^{-2}$
		and (b) $S=2$ $\textrm{mS\ensuremath{\cdot}cm}^{-2}$. Each neuron
		is driven by a constant input $I^{\textrm{input}}\sim U(5.9,6.3)$.
		Time steps and colors are the same as that in Fig \ref{fig:Hopf}.
		\label{fig:Hopf_100}}
\end{figure}

We now further check whether the library method can still capture
this Hopf phenomena in a large-scale network. We use an all-to-all
connected network with 100 excitatory neurons. Each neuron is driven
by a constant input $I^{\textrm{input}}$ that follows a uniform distribution
with mean value around the critical value. As shown in Fig \ref{fig:Hopf_100},
the library method can still capture the spike events well, with few
spikes missed because of the same coarse reason. 

\section{Results}
\label{sec: Numerical result}

\subsection{Lyapunov exponent\label{subsec:Lyapunov-exponent}}

In this section, we show that the library method with large time steps
can capture the statistical properties of the HH network. For the
sake of simplicity, we show the numerical results mainly using an
all-to-all connected network of 100 excitatory neurons with fixed
feedforward Poisson input $\nu=100$ Hz and $f=0.10$ $\textrm{ mS\ensuremath{\cdot}cm}^{-2}$.
Then the coupling strength $S$ is the only remaining variable. Other
types of HH network and other dynamic regimes can be easily extended
and similar results can be obtained. 

We first study the dynamic properties of the system by computing the
largest Lyapunov exponent which is one of the most important tools
to characterize chaotic dynamics \cite{oseledec1968multiplicative}.
The spectrum of Lyapunov exponents can measure the average rate of
divergence or convergence of the reference and the initially perturbed
orbits \cite{ott2002chaos,thompson2002nonlinear,parker2012practical}.
If the largest Lyapunov exponent is positive, then the reference and
perturbed orbits will exponentially diverge and the dynamics is chaotic,
otherwise, the dynamics is non-chaotic. 

When calculating the largest Lyapunov exponent, denoted by $\lambda$,
we use $\mathbf{X}=[X_{1},X_{2},...,X_{N}]$ to represent all the
variables of the neurons in the HH model. Denote the reference and
perturbed trajectories by $\mathbf{X}(t)$ and $\mathbf{\tilde{X}}(t)$,
respectively, then 

\begin{equation}
\lambda=\lim_{t\rightarrow\infty}\lim_{\epsilon\rightarrow0}\frac{1}{t}\ln\frac{||\mathbf{\tilde{X}}(t)-\mathbf{X}(t)||}{||\epsilon||}\label{eq:LE}
\end{equation}
where $\epsilon$ is the initial separation. However we cannot use
Eq. (\ref{eq:LE}) to compute $\lambda$ directly, because for a chaotic
system the separation $||\mathbf{\tilde{X}}(t)-\mathbf{X}(t)||$ is
unbounded as $t\rightarrow\infty$ and a numerical ill-condition will
happen. The standard algorithm to compute the largest Lyapunov exponent
can be found in \cite{parker2012practical,zhou2010spectrum,wolf1985determining}.
The regular method can compute $\lambda$ using these algorithms directly.
However, for the library method, the information of $V,m,h,n$ are
blank during the stiff period and these algorithms do not work. The
extended algorithm in \cite{zhou2009network} can solve this problem
and we use it in this Letter.

\begin{figure}[H]
	\begin{centering}
		\includegraphics[width=1\textwidth]{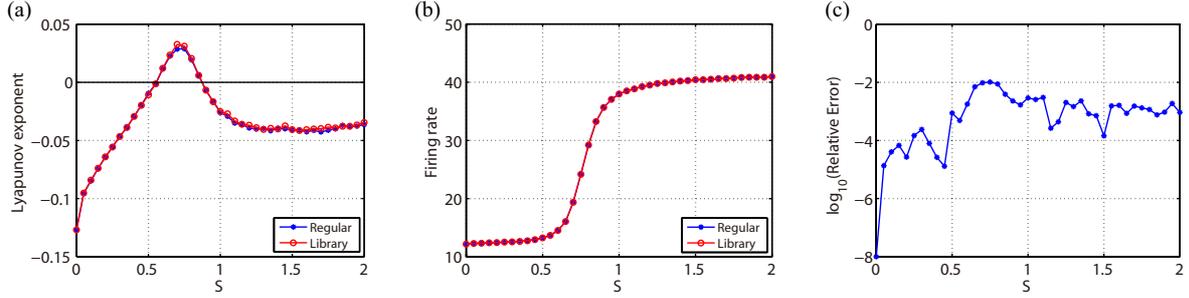}
		\par\end{centering}
	\caption{(a) The largest Lyapunov exponent of the HH network versus the coupling
		strength $S$. (b) Mean firing rate (Hz) versus the coupling strength
		$S$. (c) The relative error in the mean firing rate between the library
		and the regular method . The blue and red curves in (a) and (b) indicate the regular method with $\Delta t=2^{-5}=0.03125$ ms and library method with $\Delta t=0.25$ ms, respectively.  The total run time is 60 seconds to obtain convergent results. \label{fig:MLE}}
\end{figure}

As shown in Fig \ref{fig:MLE}(a), we compute the largest Lyapunov
exponent as a function of coupling strength $S$ from 0 to 2 $\textrm{mS\ensuremath{\cdot}cm}^{-2}$
by regular and library methods, respectively. The total run time $T$
is 60 seconds which is sufficiently long to have convergent results.
The library method with a large time step $\Delta t=0.25$ ms can
obtain accurate largest Lyapunov exponent compared with the regular
method with a small time step $\Delta t=0.03125$ ms. The results
show three typical dynamical regimes that the system is chaotic in
$0.55\lesssim S\lesssim0.875$ $\textrm{mS\ensuremath{\cdot}cm}^{-2}$
and non-chaotic in $0\lesssim S\lesssim0.55$ and $0.875\lesssim S\lesssim2$
$\textrm{mS\ensuremath{\cdot}cm}^{-2}$, depending on whether $\lambda$
is positive or negative. 

As shown in Fig \ref{fig:MLE}(b), we compute the mean firing rates,
denoted by $R$, obtained by the regular and library methods to further
demonstrate how accurate the library method is. We also give the relative
error in the mean firing rate, which is defined by

\begin{equation}
E_{\textrm{R}}^{\textrm{}}=|R_{\textrm{library}}-R_{\textrm{regular}}|/R_{\textrm{regular}}
\end{equation}
As shown in Fig \ref{fig:MLE}(c), the library method can achieve
at least 2 digits of accuracy using large time steps ($\Delta t=0.25$
ms). 

\begin{figure}[H]
	\centering{}
	\includegraphics[width=1\textwidth]{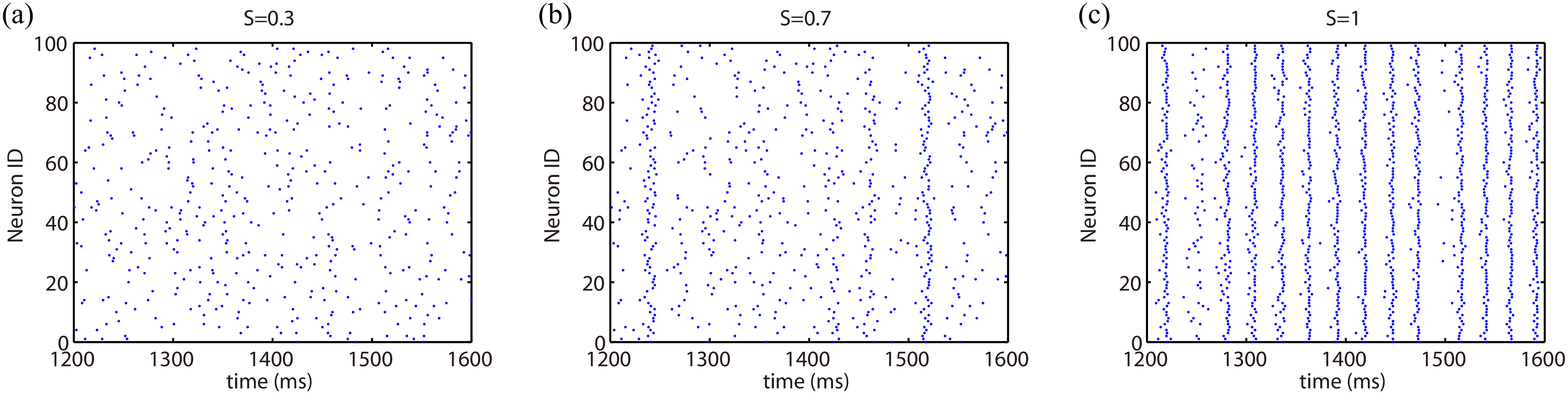}
	\caption{Raster plots of firing events in three typical dynamical regimes with
		coupling strength (a) $S=0.3$ $\textrm{mS\ensuremath{\cdot}cm}^{-2}$,
		(b) $S=0.7$ $\textrm{mS\ensuremath{\cdot}cm}^{-2}$, (c) $S=1$ $\textrm{mS\ensuremath{\cdot}cm}^{-2}$.
		Since all the three methods have similar raster, we only show the
		result obtained by the regular method. \label{fig:Raster_0.30.71.0}}
\end{figure}

From the calculation of the largest Lyapunov exponent, we have known
that there are three typical dynamical regimes in the HH model. As
shown in Fig \ref{fig:Raster_0.30.71.0}, these three regimes are
asynchronous regime in $0\lesssim S\lesssim0.55$ $\textrm{mS\ensuremath{\cdot}cm}^{-2}$,
chaotic regime in $0.55\lesssim S\lesssim0.875$ $\textrm{mS\ensuremath{\cdot}cm}^{-2}$
and synchronous regime in $0.875\lesssim S\lesssim2$ $\textrm{mS\ensuremath{\cdot}cm}^{-2}$.
Hence we choose three typical coupling strength $S=0.3,0.7$ and $1$
$\textrm{mS\ensuremath{\cdot}cm}^{-2}$ to represent these dynamical
regimes, respectively, in the following numerical tests.

\subsection{The statistical accuracy of voltage }

\begin{figure}[H]
	\centering{}
	\includegraphics[width=1\textwidth]{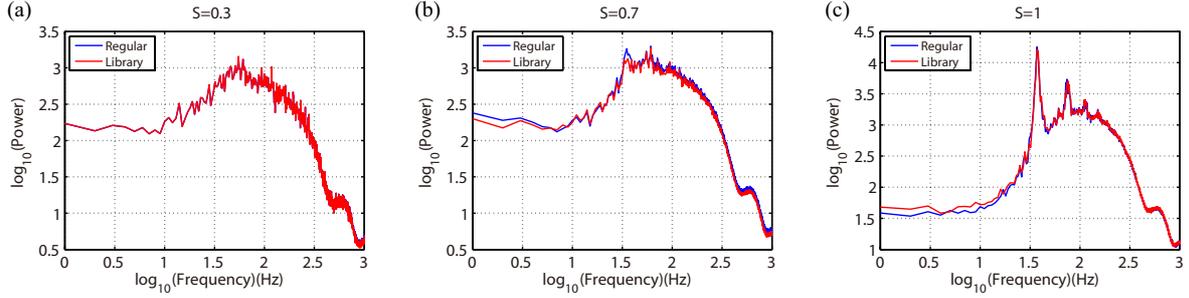}
	\caption{The averaged power spectrum of voltage trace in three typical dynamical
		regimes with coupling strength (a) $S=0.3$ $\textrm{mS\ensuremath{\cdot}cm}^{-2}$,
		(b) $S=0.7$ $\textrm{mS\ensuremath{\cdot}cm}^{-2}$, (c) $S=1$ $\textrm{mS\ensuremath{\cdot}cm}^{-2}$.
		We choose 2kHz for our sampling rate. The time steps and colors used
		are the same as the one in Fig \ref{fig:MLE}. \label{fig:fftw_0.30.71.0}}
\end{figure}

We now study the statistical accuracy of voltage. We first compute
the power spectrum of the voltage trace, averaged over all the neurons,
as shown in Fig \ref{fig:fftw_0.30.71.0}. The library method with
large time steps can capture the frequencies as well as the regular
method, $i.e.$, the library method can capture the first order information
of the voltage. 

We should point out that when computing the power spectrum, we need
the trace of voltage $V$, which is blank during the stiff period
in the library method. To solve this problem, we record the trace
of voltage $V(t;I^{\textrm{th}},m^{\textrm{th}},h^{\textrm{th}},n^{\textrm{th}})$
when building the library, where $I^{\textrm{th}},m^{\textrm{th}},h^{\textrm{th}},n^{\textrm{th}}$
are the corresponding initial values. In this Letter, we use sampling
rate $256$ kHz to record. Then we can perform a linear interpolation
to estimate the voltage during the stiff period in the library method.
Note that the record process of the trace of voltage is not necessary
to evolve the HH model with the library method. 

\begin{figure}[H]
	\begin{centering}
		\includegraphics[width=1\textwidth]{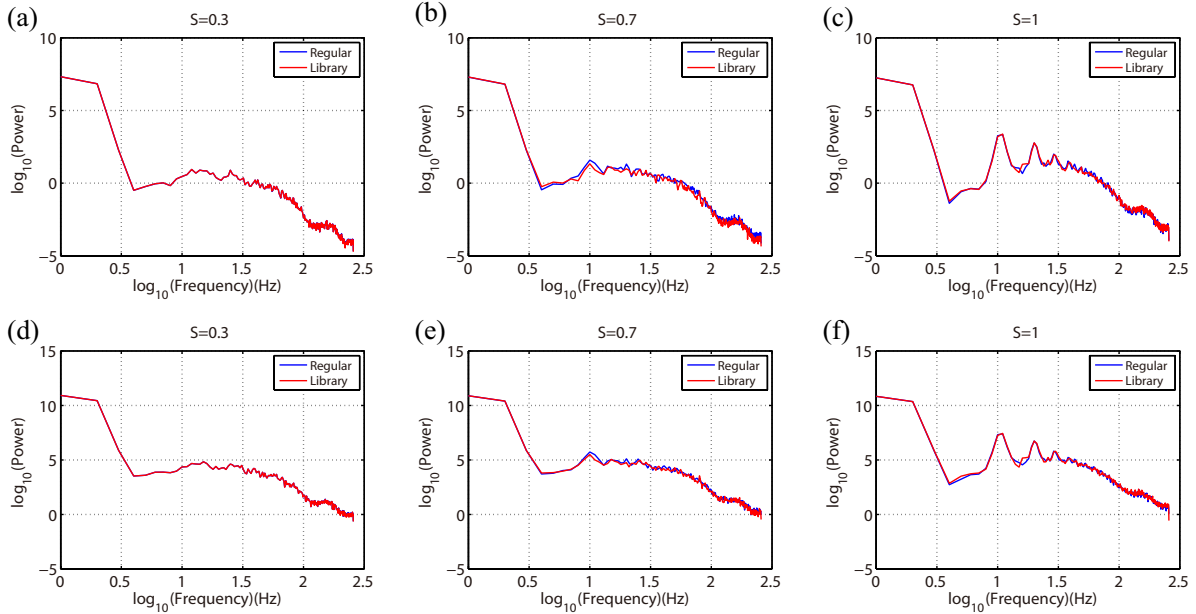}
		\par\end{centering}
	\caption{Top panel : The $L^{2}$-norm of cpsd between two randomly chosen
		neurons with coupling strength $S=0.3,0.7$ and $1$ $\textrm{mS\ensuremath{\cdot}cm}^{-2}$.
		Bottom panel : The $L^{2}$-norm of cpsd among three randomly chosen
		neurons with coupling strength $S=0.3,0.7$ and $1$ $\textrm{mS\ensuremath{\cdot}cm}^{-2}$.
		For example, given the traces of voltage $V_{1}(t),V_{2}(t)$ and
		$V_{3}(t)$, we use the information of $V_{1}(t),V_{2}(t+\tau)$ and
		$V_{3}(t)$ to compute cpsd. The time steps and colors used are the
		same as the one in Fig \ref{fig:MLE}. \label{fig:cpsd_2_3}}
\end{figure}

We further demonstrate that the library method can capture higher
order information of voltage, $e.g.$, the second and third order.
Given the voltage traces of two neurons, we can compute their cross
power spectral density (cpsd) and its $L^{2}$-norm. We use cpsd function
in Matlab to compute in this Letter. Fig \ref{fig:cpsd_2_3}(a-c)
show the $L^{2}$-norm of cpsd between two randomly chosen neurons
with different coupling strength $S=0.3,0.7$ and $1$ $\textrm{mS\ensuremath{\cdot}cm}^{-2}$,
while Fig \ref{fig:cpsd_2_3}(d-f) show the results among three
neurons. Therefore, the library method with large time steps can also
capture the second and third order information of voltage as well
as the regular method. 

\subsection{The statistical accuracy of spikes }

Thanks to the advances in the spike train measurement like multiple
electrode recording techniques, neuroscientists can obtain large amounts
of spike data much easier than the voltage. The data can be used in
statistical tools, $e.g.$, transfer entropy, maximum entropy and
Granger causality \cite{perkel1967neuronal,shlens2006structure,quinn2011estimating,zhou2014granger,xu2017dynamical}.
Based on the spike trains, these tools can not only solve the directed
causal information and network inference problem \cite{ito2011extending,zhou2014granger}
but also probe the structure of fire patterns \cite{shlens2006structure,shlens2009structure,xu2017dynamical}.
Another advantage over the voltage is that the spike train data is
binary, hence the state space is very small. For example, considering
the vector $\mathbf{x}^{(10)}=(x_{1},x_{2},...,x_{10})$ from the
spike data, the state space is only $2^{10}=1024$. So it is much
easier applied to these tools while achieving faster calculation.
Therefore, it is necessary to check if the library method can still
obtain accurate spike trains. 

\begin{figure}[H]
	\begin{centering}
		\includegraphics[width=1\textwidth]{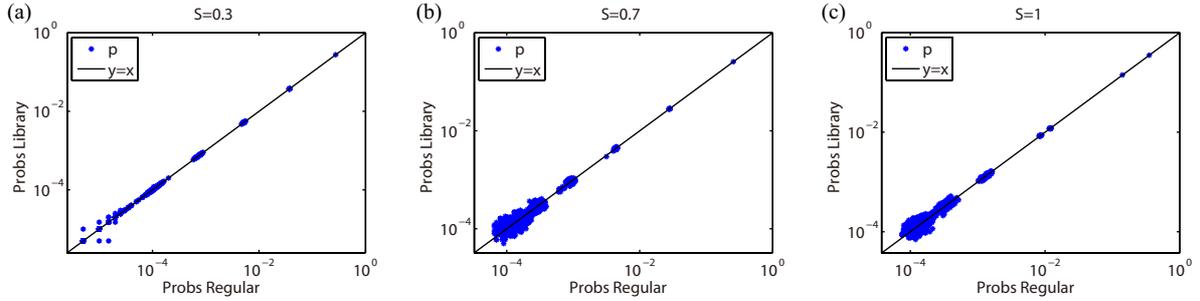}
		\par\end{centering}
	\caption{The comparison of the probabilities of the 10-dimesional vector by
		regular and library method. The stars indicate the probability of
		the same vector computed by two methods. Therefore, the stars on the
		diagonal line $y=x$ mean the probabilities are the same and the compared
		method is as well as the regular method. Time steps are the same as
		the one in Fig \ref{fig:MLE}. Total run time is $10000$ seconds
		to obtain precise distribution. \label{fig:fire_pattern_10}}
\end{figure}

As shown in Fig \ref{fig:MLE}, we demonstrate the accuracy of
mean firing rate by the library method. However, this is the first
order information of spike trains and the at least two digits of accuracy
is still very coarse. We now further demonstrate the statistical accuracy
of spikes. When evolving the HH model (\ref{eq: HH}), we randomly
choose 10 neurons from the network, record their spike times and transform
them into binary time series with a time bin 10 ms. We set the value
1 if there is a spike event during the time bin and 0 otherwise. Therefore,
the 10 neurons can make up a 10-dimensional vector with total 1024
kinds of combinations. After evolving with a sufficiently long run
time, we can obtain a quite accurate distribution of the 10-dimensional
vector. As shown in Fig \ref{fig:fire_pattern_10}, we compare
the probabilities of the 10-dimensional vector computed by the library
and regular methods. Each star indicates the probability of the same
vector (fire pattern) computed by the two methods. If the stars are
on the diagonal line, then the library method can capture quite the
same distribution as the regular method. 

We also do chi-square two sample tests for the comparison of distributions
between the library and regular methods. The test statistic with different
coupling strength all has p-value greater than $1-10^{-10}$. Therefore,
the distributions from the regular and library methods cannot be distinguished
in a statistical sense, $e.g.$, the library method can capture the
fire patterns very well. Since the statistical tools need only the
distribution of fire patterns, we can use the spike trains computed
by the library method with large time steps in application.

\subsection{Computational efficiency}

\begin{figure}[H]
	\centering{}
	\includegraphics[width=1\textwidth]{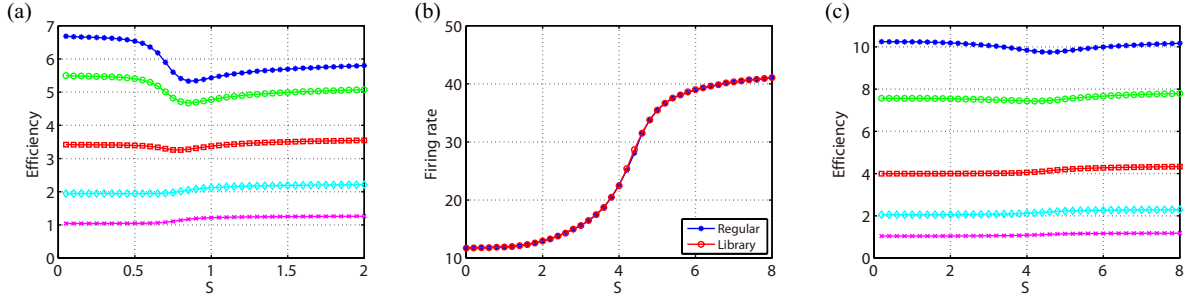}
	\caption{(a) The efficiency of the library method versus the coupling strength
		$S$ with the time steps $\Delta t=0.354,0.25,0.125,0.0625,0.03125$
		ms from top to bottom. Total run time is 50 seconds. (b) Mean firing
		rate (Hz) and (c) efficiency in a sparse network with 100 excitatory
		neurons randomly connected with probability 15\%. The coupling strength
		$S_{ij}=S/N$ if there is a connection from $j$-th neuron to $i$-th
		neuron. Other parameters and time steps are the same as that in (a)
		\label{fig:efficiency}}
\end{figure}

We now demonstrate the computational efficiency of the library method
by comparing the time each method costs with the same total run time.
To reduce the system error from the computer, we use a sufficiently
long run time of 50 seconds. In the comparison, we fix the time step
$\Delta t=2^{-5}=0.03125$ ms in the regular method and change time
step in the library method with a maximum value $\Delta t=0.354$
ms. As shown in Fig \ref{fig:efficiency}(a), the library method
can overall stably obtain a maximum computational speedup around 6
times compared with the regular method. We find that the computational
speedup is not increased linearly with the time step. This is because
when we use a large time step, the spike-spike correction procedure
requires more computation. Besides, once a neuron fires a spike, the
library method should call the library for once and evolve the parameters
$G$ and $H$ during the stiff period as usual which also costs some
time. Therefore, the computational speedup is not increased straightforwardly
with the time step. 

In the all-to-all connected network, once a neuron fires a spike,
all the other neurons should update their state and preliminarily
evolve the HH equations (\ref{eq: HH}) to find the next spike time.
We should point out that real neuron networks are often sparsely connected
\cite{golomb2000number,song2005highly,honey2007network,gerstner2014neuronal}.
So only a few portion of the remaining neurons should update their
state when there is a spike event. Therefore, the efficiency shown
in Fig \ref{fig:efficiency}(a) is underestimated. We consider
a sparse network with 100 excitatory neurons randomly connected with
probability 15\%. As shown in Fig \ref{fig:efficiency}(c), the
library method can achieve at most 10 times of efficiency with the
maximum time step $\Delta t=0.354$ ms. 

\subsection{Extension}

\subsubsection{More realistic networks}

The results presented in section \ref{sec: Numerical result} are
mainly based on an all-to-all coupled network. We now consider networks
with more complicated structure, $e.g.$, let the firing rates and
coupling strength of the model neurons have a distribution \cite{song2005highly,hromadka2008sparse}
rather than a homogeneously value of $S/N$ or nearly fixed firing
rate. And further check the validity of the library method in more
complicated situations.

\begin{figure}[H]
	\begin{centering}
		\includegraphics[width=1\textwidth]{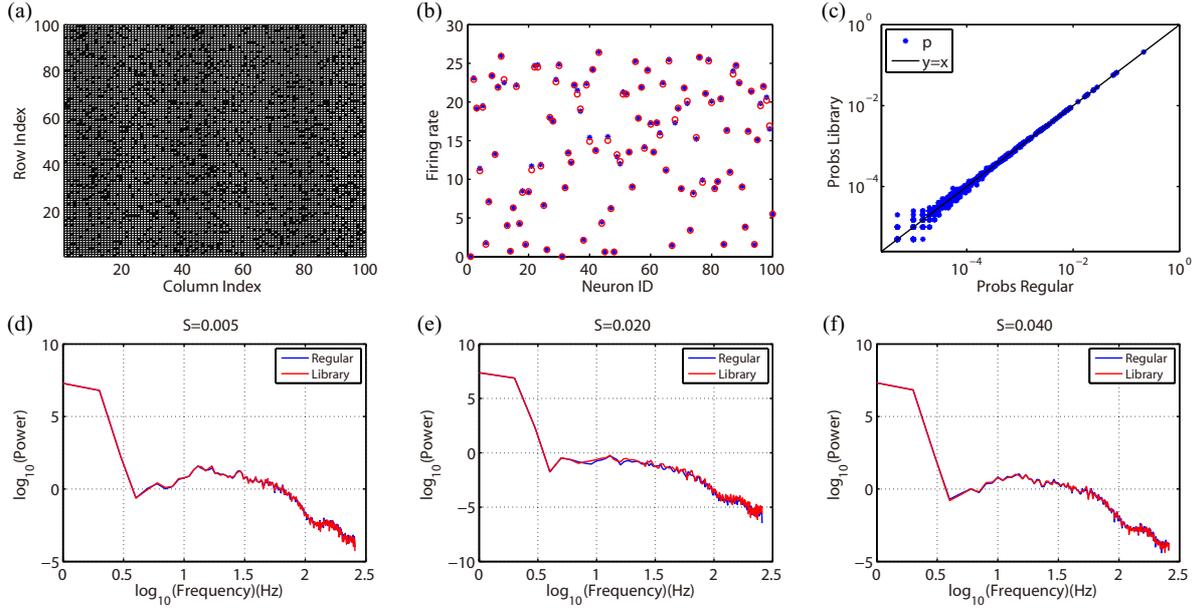}
		\par\end{centering}
	\caption{The performance of library method in a network of 100 excitatory neurons,
		randomly coupled with probability 15\%. The coupling strength $S_{ij}\sim U(0,0.04)$
		$\textrm{ mS\ensuremath{\cdot}cm}^{-2}$, the Poisson input rate $\nu_{i}\sim U(0,200)$
		Hz, the Poisson input strength $f=0.1$ $\textrm{ mS\ensuremath{\cdot}cm}^{-2}$.
		The blue and red colors indicate the result from regular method with
		$\Delta t=2^{-5}=0.03125$ ms and library method with $\Delta t=0.25$
		ms, respectively. (a) The synaptic adjacency matrix with a black color
		indicating a connection and white color otherwise. (b) Firing rate
		of each neuron. (c) The comparison of the probabilities of the 10-dimesional
		vector. The $L^{2}$-norm of cpsd between two neurons with different
		coupling strength (d) $S=0.005$ $\textrm{\ensuremath{\textrm{ mS\ensuremath{\cdot}cm}^{-2}}}$,
		(e) $S=0.020$ $\textrm{ mS\ensuremath{\cdot}cm}^{-2}$, (f) $S=0.040$
		$\textrm{ mS\ensuremath{\cdot}cm}^{-2}$ \label{fig:UUp=0.15s=2u=0.1}}
\end{figure}

We first use a network of 100 excitatory neurons, randomly coupled
with probability 15\% as shown in Fig \ref{fig:UUp=0.15s=2u=0.1}(a).
The coupling strength of each connection follows a Uniform distribution
$U(0,0.04)$ $\textrm{ mS\ensuremath{\cdot}cm}^{-2}$ and the Poisson
input rate to each neuron follows also a Uniform distribution $U(0,200)$
Hz. Then the firing rate of each neuron has a large range from 0 to
tens of Hz as shown in Fig \ref{fig:UUp=0.15s=2u=0.1}(b).
We check the statistical accuracy of both voltage and fire patterns
in Fig \ref{fig:UUp=0.15s=2u=0.1}(c-f). The
test statistic of chi-square two sample tests always has $p$-value
greater than $1-10^{-10}$, unless stated otherwise. The library method
with a large time step $\Delta t=0.25$ ms can still achieve good
performance compared with the regular method. 

\begin{figure}[H]
	\begin{centering}
		\includegraphics[width=1\textwidth]{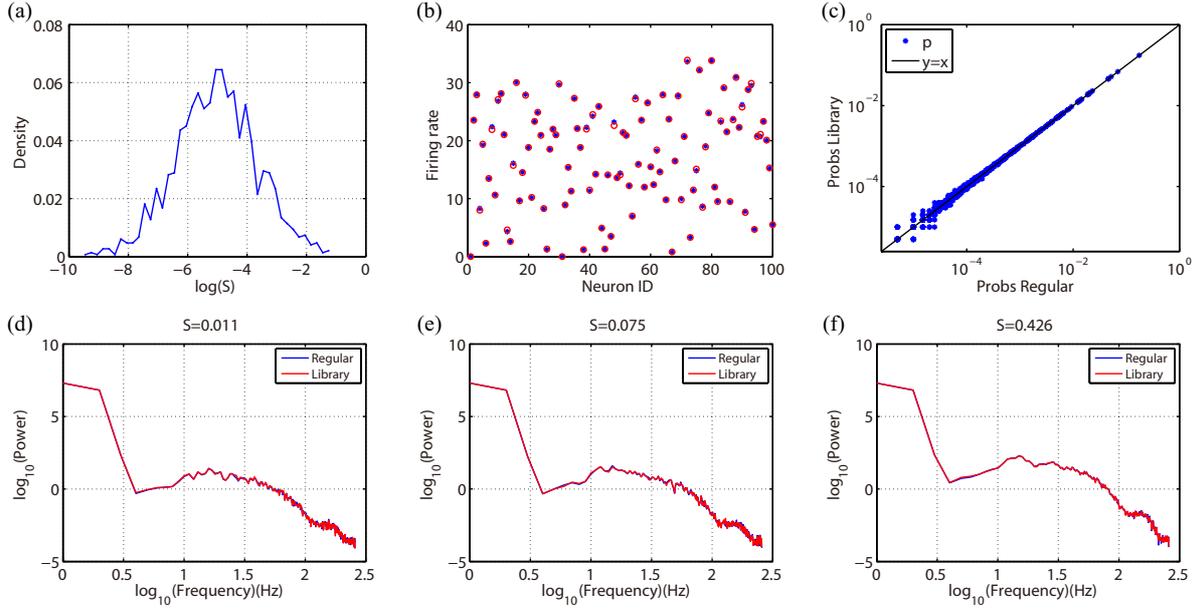}
		\par\end{centering}
	\caption{The performance of library method in the same network as shown in
		Fig \ref{fig:UUp=0.15s=2u=0.1} but with coupling
		strength $S_{ij}\sim\textrm{Lognormal}(-4.890,1.956)$ $\textrm{ mS\ensuremath{\cdot}cm}^{-2}$
		(the mean value of $S_{ij}=0.02$ $\textrm{ mS\ensuremath{\cdot}cm}^{-2}$).
		The colors and time steps are the same as that in Fig \ref{fig:UUp=0.15s=2u=0.1}.
		(a) The distribution of the coupling strength. (b) Firing rate of
		each neuron. (c) The comparison of the probabilities of the 10-dimesional
		vector. The $L^{2}$-norm of cpsd between two neurons with different
		coupling strength (d) $S=0.011$ $\textrm{ mS\ensuremath{\cdot}cm}^{-2}$,
		(e) $S=0.075$ $\textrm{ mS\ensuremath{\cdot}cm}^{-2}$, (f) $S=0.426$
		$\textrm{ mS\ensuremath{\cdot}cm}^{-2}$. \label{fig:LNUp=0.15s=2u=0.1}}
\end{figure}

According to data from living cortical neuronal networks, the coupling
strength follows a Log-normal distribution \cite{song2005highly}.
We then adjust the coupling strength following from a Uniform distribution
to a Log-normal distribution but keep the mean value of 0.02 $\textrm{ mS\ensuremath{\cdot}cm}^{-2}$.
As shown in Fig \ref{fig:LNUp=0.15s=2u=0.1},
the library method can still achieve good performance. 

\subsubsection{Extended HH-type models}

In this part, we apply the library method to other types of HH model
for the four most prominent classes of neurons. They are ``regular
spiking'' (RS), \textquotedblleft fast spiking\textquotedblright{}
(FS), \textquotedblleft intrinsically bursting\textquotedblright{}
(IB) and \textquotedblleft low-threshold spike\textquotedblright{}
(LTS) cells according to the pattern of spiking and bursting in intracellular
recordings \cite{connors1990intrinsic}. These HH-type models are
obtained by fit method based on different experimental data like from
rat somatosensory cortex in vitro, ferret visual cortex in vitro,
cat visual cortex in vivo and cat association cortex in vivo. All
the four extended HH-type models can be described by the following
equation \cite{pospischil2008minimal}: 

\begin{equation}
C_{m}\frac{dV_{i}}{dt}=-g_{\textrm{leak}}(V_{i}-E_{\textrm{leak}})-I_{i}^{\textrm{Na}}-I_{i}^{\textrm{Kd}}-I_{i}^{\textrm{M}}-I_{i}^{\textrm{T}}-I_{i}^{\textrm{L}}+I_{i}^{\textrm{input}}
\end{equation}
where $V_{i}$ is the membrane potential of the $i$-th neuron, $I_{i}^{\textrm{Na}},I_{i}^{\textrm{Kd}},I_{i}^{\textrm{M}},I_{i}^{\textrm{T}}$
and $I_{i}^{\textrm{L}}$ are voltage-dependent currents, $I_{i}^{\textrm{input}}$
is the input current, $C_{m}=1\mu\textrm{F\ensuremath{\cdot}c\ensuremath{m^{-2}}}$
is the specific capacitance of the membrane, $g_{\textrm{leak}}$
and $E_{\textrm{leak}}$ are the resting membrane conductance and
reversal potential, respectively. Detailed functions and parameters
for the four extended HH-type models \cite{pospischil2008minimal}
are given in Appendix.

\begin{figure}[H]
	\begin{centering}
		\includegraphics[width=1\textwidth]{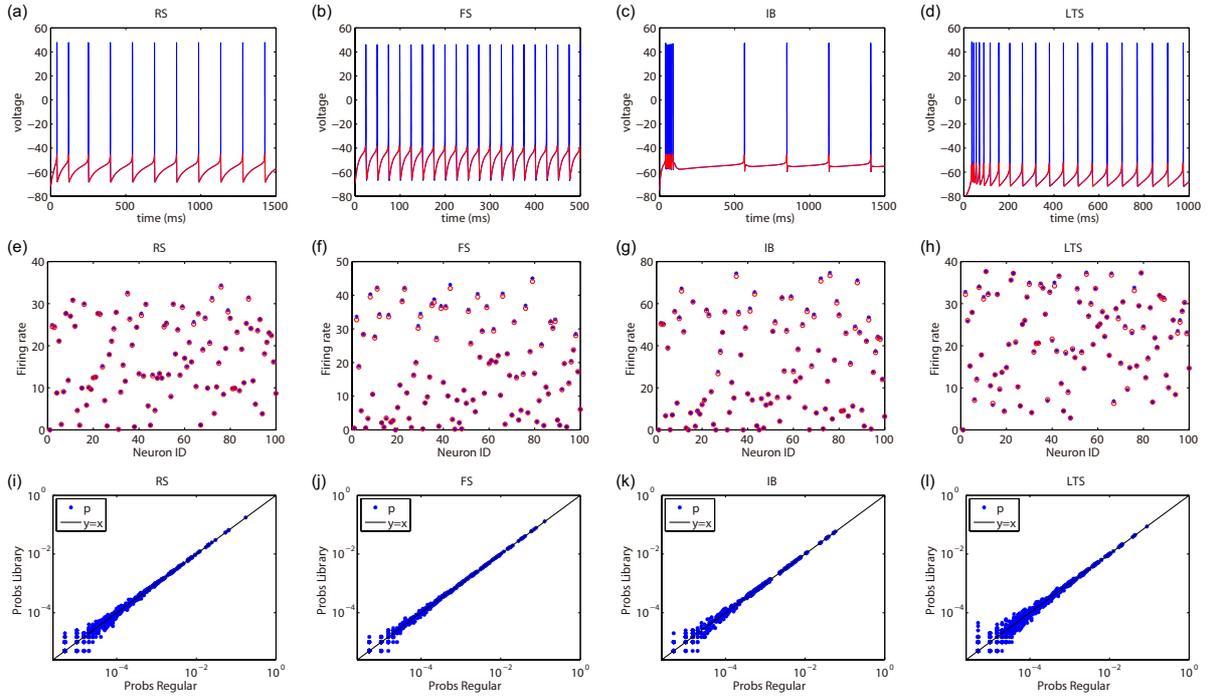}
		\par\end{centering}
	\caption{The validity of library method for the extended HH-type models. Top
		panel: Voltage traces of a single neuron driven by a constant input.
		Constant input are $I^{\textrm{input}}=0.8,4,0.4,2.2$ $\mu\textrm{A\ensuremath{\cdot}cm}^{-2}$
		for the RS, FS, IB and LTS neurons, respectively. Middle panel: Firing
		rate of each neuron of the RS, FS, IB and LTS networks. Bottom panel:
		The comparison of the probabilities of the 10-dimesional vector for
		the RS, FS, IB and LTS networks. The colors and time steps are the
		same as that in Fig \ref{fig:UUp=0.15s=2u=0.1}.
		All the four networks have 100 neurons (RS and IB are excitatory network
		while FS and LTS are inhibitory network), coupled as in Fig \ref{fig:UUp=0.15s=2u=0.1}(a),
		Poisson input strength $f=0.1$ $\textrm{ mS\ensuremath{\cdot}cm}^{-2}$.
		Other parameters are RS: coupling strength $S_{ij}\sim\textrm{Lognormal}(-5.757,2.303)$,
		Poisson input rate $\nu_{i}\sim U(0,200)$ Hz; FS: $S_{ij}\sim\textrm{Lognormal}(-5.757,2.303)$,
		$\nu_{i}\sim U(200,600)$ Hz; IB: $S_{ij}\sim\textrm{Lognormal}(-6.623,2.649)$,
		$\nu_{i}\sim U(0,100)$ Hz; LTS: $S_{ij}\sim\textrm{Lognormal}(-5.757,2.303)$,
		$\nu_{i}\sim U(0,400)$ Hz.\label{fig:4type_LNU}}
\end{figure}

RS neurons are the most typical neurons in neocortex and is in general
excitatory. When injected by a constant depolarizing current, the
neurons can fire with short inter-spike-interval (ISI) at first and
then the ISI increases and tends to be stable as shown in Fig \ref{fig:4type_LNU}(a).
This is called spike-frequency adaptation which is one of the mean
features. We also use an excitatory RS-type network with coupling
strength following a Log-normal distribution, Poisson input rate following
a Uniform distribution to illustrate the validity of library method.
The statistical accuracy of spikes are given in Fig \ref{fig:4type_LNU}(e,
i). 

We should point out that the extended HH-type model contains more
voltage-dependent currents, $e.g.$, the IB-type model requires 7
parameters: $I^{\textrm{th}},m^{\textrm{th}},h^{\textrm{th}},n^{\textrm{th}},p^{\textrm{th}},q^{\textrm{th}},r^{\textrm{th}}$
when building the library. As a consequence, the trace of voltage
during the stiff period will occupy huge storage space (>10 gigabyte
in binary form). Therefore, we do not record the trace of voltage
when building the library and present the accuracy of voltage. 

FS neurons are another kind of typical neurons in cortex and is in
general inhibitory. The mean feature of a FS neuron is that it can
fire high-frequency spikes with little or no adaptation, as shown
in Fig \ref{fig:4type_LNU}(b). We use an inhibitory FS-type network
to show the validity of library method in Fig \ref{fig:4type_LNU}(f,
j). 

IB neurons are usually excitatory and can produce bursts of spikes,
while LTS neurons are usually inhibitory and can fire high-frequency
spikes with spike-frequency adaptation. The validity of library method
for the IB and LTS network are given in the third and last column
of Fig \ref{fig:4type_LNU}, respectively.

\section{Conclusion}

In conclusion, we have shown a modified library method to deal with
the stiff part during the firing event in evolving the HH model. The
library method can enlarge time step (maximum time step 0.354 ms)
to reduce the computational cost while achieving high accurate statistical
information of the HH neurons. It is worthwhile pointing out that
the library method with large time steps can capture the fire patterns
or the distribution of the spikes as well as the regular method. This
holds a spectral attraction to apply to statistical analysis like
the transfer entropy and maximum entropy which require a sufficiently
long run time $\sim20$ minutes to obtain a precise distribution of
spikes. However, due to the extra error introduced by calling the
library, it can never obtain numerical convergence. But the library
method can still retain most of the properties of HH neurons like
the chaotic dynamics which is observed in the HH model by computing
the largest Lyapunov exponent.

We emphasize that the library method is very attractive with a stably
maximum 10 times of speedup in a sparse network. The remarkable speedup
of library method holds in spite of the size of the network or the
structure of connectivity. 

We can successfully extend the idea of library method in more complicated
HH-type models with bursting and adaptation behavior. The way to build
and use the library are the same as that in the standard HH model
although there are more voltage-dependent currents. The library method
can also capture most of the statistical properties of these HH-type
neurons with high times of speedup. 

Finally, we emphasize that the spike-spike correction procedure \cite{rangan2007fast}
is necessary in the two methods, especially using a large time step
in the library method. This procedure ensures that the spiking sequences
estimated are accurate. Even in a strongly coupled network, the synaptic
interactions are still correct and will not influence the accuracy
of the library method.

\bibliographystyle{unsrt}
\bibliography{reference}

%\begin{small}
%\bibliographystyle{plain}
%\bibliography{ETDrefer}
%\end{small}

%%%%%%%%%%%%%%%%%%%%%%%%%%%%%%%%%%
\end{document}